\documentstyle[12pt,aaspp4]{article}
\lefthead{Dolphin}
\righthead{WFPC2 Calibration}
\begin{document}

\title{The Charge Transfer Efficiency and Calibration of WFPC2}

\author{Andrew E. Dolphin}
\affil{National Optical Astronomy Observatories, P.O. Box 26372, Tucson, AZ 85726\\
Electronic mail: dolphin@noao.edu}

\begin{abstract}
A new determination of WFPC2 photometric corrections is presented, using
HSTphot reduction of the WFPC2 Omega Centauri and NGC 2419 observations
from January 1994 through March 2000 and a comparison with ground-based
photometry.  No evidence is seen for any position-independent photometric
offsets (the ``long-short anomaly''); all systematic errors appear to be
corrected with the CTE and zero point solution.  The CTE loss time dependence
is determined to be very significant in the Y direction, causing
time-independent CTE solutions (Stetson 1998; Saha, Lambert, \& Prosser
2000) to be valid only for a small range of times.  On average, the present
solution produces corrections similar to Whitmore, Heyer, \& Casertano
(1999), although with an improved functional form that produces less scatter
in the residuals and determined with roughly a year of additional data.  In
addition to the CTE loss characterization, zero point corrections are also
determined as functions of chip, gain, filter, and temperature.  Of interest,
there are chip-to-chip differences of order 0.01-0.02 magnitudes relative
to the Holtzman et al. (1995) calibrations, and the present study provides
empirical zero point determinations for the non-standard filters such as
the frequently-used F450W, F606W, and F702W.
\end{abstract}


\section{Introduction}

Shortly after the installation of WFPC2 on Hubble Space Telescope, it was
discovered that the camera suffered from a charge transfer inefficiency,
causing stars at the top of each chip to lose 10-15\% of their charge
while being read out.  This effect was significantly reduced, although not
eliminated, by cooling the camera from $-76^{\circ}$C to $-88^{\circ}$C.
Holtzman et al. (1995, hereafter H95) gave initial estimates of the charge
loss, and observed that it appeared to be related to the background level.
Whitmore \& Heyer (1997) quantified the background dependence, and also
detected a count dependence.  Furthermore, a charge transfer loss was also
seen in the X direction, which was also characterized by Whitmore \& Heyer
(1997).

A further complication arose when it was discovered that the Y and possibly
X losses were growing with time, a dependence characterized by Whitmore
(1998) and most likely caused by radiation damage. The most recent paper on
this topic from the STScI WFPC2 group is Whitmore, Heyer, \& Casertano (1999,
hereafter WHC99), who combine the two earlier results and have a longer time
baseline (through February 1999) with which to determine the time
dependence.

In addition to these studies, there have been two recent independent
determinations of the CTE loss.  Stetson (1998, hereafter S98), using
DAOPHOT reduction of the Omega Cen and NGC 2419 calibration fields, derived
a calibration that produced similar results to Whitmore \& Heyer (1997)
but finding no significant time dependence (+0.0012 magnitudes per year
for the typical star).  Saha, Lambert, \& Prosser (2000, hereafter SLP00),
in a paper detailing errors in DoPHOT Cepheid reductions, found significantly
different results from previous studies in their reductions of the NGC 2419
field: no detectable XCTE loss and no count dependence of the YCTE.  The
presence of very different conclusions as to the dependencies of the CTE
loss leads to the uncomfortable possibility that the amount of the CTE loss
is package-dependent, as aperture photometry, DAOPHOT, and DoPHOT reductions
have produced different results.  As one would hope that the stellar
brightnesses scale linearly from package to package, the most likely
cause of such a dependence would be in the background calculation.  DoPHOT
backgrounds, especially, contain a significant amount of starlight and
thus it is possible that the lack of an obvious count dependence in the
DoPHOT CTE solution was caused by the presence of a count dependence in
the background level.

In addition to the CTE effect, a position-independent charge loss was
detected and dubbed the ``long-short anomaly'' because it was first seen
as a difference in the magnitudes of a star in short and long exposures.
Casertano \& Mutchler (1998, hereafter CM98) determined this effect to be a
function of the number of counts rather than the exposure time through an
analysis of the NGC 2419 calibration field, and determined a correction
formula.  SLP00 found evidence of such a position-independent correction
in their F814W observations of the same data, but not in the F555W
observations.  S98 found no evidence of a position-independent anomaly in
any of his data, despite also using the NGC 2419 calibration data.

Finally, the validity of the H95 calibrations have been called into
question by a number of studies.  In addition to the ``long'' zero points
measured by Kelson et al. (1996), Saha et al. (1996), Hill et al. (1998),
and others, which were all about 0.05 magnitudes fainter than the H95
zero points, S98 and SLP00 have determined zero point corrections using
their CTE solutions.

This paper uses HSTphot (Dolphin 2000) reductions of the Omega Centauri
and NGC 2419 standard fields in an attempt to address the unresolved
issues from the studies mentioned above.  At the very least, should the
CTE loss be determined to be package-dependent, it will be necessary to
provide a solution that is valid for HSTphot reductions.  However, a
unified explanation of the different effects is preferred and will be
sought.  This study also provides something of a ``trial by fire'' for
HSTphot, given the large amount of data (over 1000) images that were
reduced, primarily non-interactively.

\section{Observations and Reduction}

As with S98, this study was based on observations of the Omega Cen and
NGC 2419 calibration fields.  The Omega Cen data comprise the bulk of the
observations and span a wide baseline of epoch (January 1994 through
March 2000), use all of the non-UV filters, and contain observations at
both gain settings and both temperatures.  However, nearly all of these
data contain little or no background, so the NGC 2419 field (also used by
S98 and SLP00) was added to improve the background baseline.

\subsection{Ground-based Data}

The ground-based Omega Cen data were those of Walker (1994), with stars
fainter than V=21.0 eliminated because of the large scatter in the data at
the faint end.  The ground-based NGC 2419 data were provided by Peter
Stetson, in which a similar faint-end cut had already been made.  The
faint cuts in both data sets, in addition to reducing the amount of
low-quality data in the fits, avoid the photometry bias that is found just
above the cutoff.  As the WFPC2 data are generally deeper than the
ground-based data, they also avoid this effect.

Given the superior resolution of HST over ground-based telescopes, it is
not surprising that many of the ground-based stars were resolved into
multiple star systems when observed by HST.  In order to eliminate errors
from this effect, all WFPC2 stars that fell within 0.8 arcsec of the
ground-based standard star were combined into a single star.  If the
combined magnitude was more than 0.05 magnitudes brighter than that of the
brightest constituent star, the standard star was thrown out.  Otherwise,
the star was kept, with the combined magnitude used for the WFPC2
magnitude in the analysis.  If at least 25\% of all images with detections
of a star contained bright companions, the star was eliminated from the
analysis altogether.  (In other cases, it was assumed that a cosmic ray was
responsible for the second detection, as no cosmic ray cleaning could be
made.)  In this process, 29 Omega Cen stars and 13 NGC 2419 stars were
eliminated.

For the remaining stars, the expected WFPC2 flight system magnitudes
were calculated using the H95 zero points and color terms with the equation
\begin{equation}
\hbox{WFPC2}=\hbox{SMAG}+Z_{FG}-Z{FS}-T_{1,FS}\hbox{SCOL}-T_{2,FS}\hbox{SCOL}^2,
\end{equation}
where the values are defined as in H95.  For this analysis, it is assumed
that the color terms from H95 are correct, a necessary assumption given
that these data do not contain a range of colors sufficient to
re-calibrate the color terms.  However, filter-dependent corrections are
determined for the zero points.

\subsection{WFPC2 Data}

WFPC2 observations of the Omega Cen and NGC 2419 standard fields were
obtained from the STScI archive, using on-the-fly calibration to process
the images using the best available calibration data at the time of
retrieval.  The Omega Cen data consist of 795 images, mostly in the
standard filters (117 F439W, 271 F555W, 88 F675W, and 235 F814W), with
additional images reduced but discarded because of insufficient overlap
with the ground-based field.  Additionally, data in secondary filters were
also obtained (8 F380W, 8 F410M, 4 F450W, 6 F467M, 9 F547M, 7 F569W, 13
F606W, 2 F622W, 9 F702W, 7 F785LP, 2 F791W, 2 F850LP, and 4 F1042M).  U
observations, although available, were omitted because of the lack of
ground-based comparison photometry in the Walker (1994) data.

The NGC 2419 data consist of 51 images (7 F555W and 44 F814W).  F300W
images were also available for NGC 2419, but were omitted in this
analysis due to the lack of interest in calibrating F300W (as per the
recommendation of H95).  In the interest of producing the best possible
calibrations, the chips containing NGC 2419 itself (WFC4 for five of the
images; WFC2 for the remainder) were omitted from the analysis because of
the higher crowding.  This caused a mean magnitude offset of $\sim$0.02
magnitudes, which is reasonable in crowded field photometry but undesirable
in a calibration study, especially since the NGC 2419 data provide most of
the high-background points.

Photometry was made using the usual HSTphot recipe, except that cosmic ray
cleaning was omitted because the images were reduced individually.
Aperture corrections were made to correct to the 0.5 arcsec aperture of
H95.  In order to reduce the number of bad points in the analysis, all
stars were required to have $\chi \le 1.5$ and $-0.3 \le sharpness \le 0.3$
to be used in the calibration solution.  Additionally, all stars near the
WFPC2 saturation limit were removed, as were stars with fewer than 105
electrons, which contributed more noise than signal to the solution.  In
all, 843 images were used for the solution, producing a total of 58728
cleanly-fit stars that were matched to the standard stars.

\subsection{Background}

As noted in Section 1, the most likely explanation of the apparent package
dependence of the CTE loss is that different photometry packages calculate
the background differently.  Specifically, any package that determines a
background level close to the star will invariably measure the wings of the
star as part of the background.  The DoPHOT sky measurement, as made by
Saha et al. (1996) and SLP00, is perhaps the most severe example of this
effect, with its average sky pixel containing $\sim$0.2\% of the star's
total light.  HSTphot, which also determines sky values near the star
(although not as close as DoPHOT), shows this dependence at a smaller
level.

In an attempt to create as generally useful a CTE determination as
possible (as well as to provide independent parameters for the solution),
the count dependence of the background was solved and subtracted from the
background levels before running the calibration solution.  Because the
background contamination is proportionally the most significant at low
background levels, it was calculated using the low-background Omega Cen
data only.

This dependence was determined to be chip-dependent, in that it was about
an order of magnitude smaller in PC1 but the same in the three WFCs,
wavelength-dependent (smallest for B and V), and temperature-dependent
(slightly larger at cold temperature than at warm temperature).  The
correction equations and coefficients are not given here, since this result
almost certainly will not apply to other software packages.  For example,
DoPHOT background levels have a factor of $\sim20$ greater dependence on
the number of counts than do HSTphot background levels.

Finally, as a negative background is a non-physical result of readout noise
or bias errors on a very small background, all negative values were set to
zero.  The background and counts were converted to electrons by multiplying
by the gain (7 or 14), and the background then softened according to S98's
procedure (background = $\sqrt{1+{N_{electrons}}^2}$) to allow logarithms
to be taken at zero background.  For the remainder of this paper, $counts$
will refer to the number of electrons detected for a star, while
$background$ will refer to the background level in electrons, so that the
value of the gain is unimportant.  If discussing values in DN rather than
electrons, it will be made explicit.

\section{Characterization}

Before determining the CTE and zero point solution, some preliminary work
needed to be done in order to constrain the form of the solution.  It
should be stressed that the constraints determined in this section are
empirically determined from observed differences between WFPC2 and
ground-based data, rather than based on any assumed physical model of the
charge loss.

\subsection{Position-Independent Corrections}

The nature of the data in this present study - WFPC2 photometry with
comparison ground-based photometry - permits a more direct study of the
position-independent error (``long-short anomaly'') than what was given
by CM98.  The complete sample of 58728 stars was restricted to points with
limited ranges of counts, background, and epoch, thus producing a
relatively homogeneous data set that should have a common set of
correction factors.  To maximize the ability to determine the
position-independent correction, the $counts$ were restricted to between
210 and 350, for which the average position-independent correction (based
on CM98) should be about 0.23 magnitudes.  The $background$ was required
to be 3.5 or less, which would include the greatest amount of data.  The
differences between WFPC2 and ground-based magnitudes were then fit to the
formula
\begin{equation}
\Delta mag = c_0\ +\ c_1\frac{y}{800}\ +\ c_2\frac{x}{800} +\ \Delta ZP.
\end{equation}
($\Delta counts$ could have been used equally well, as the next section
demonstrates, with the same result produced.)  $\Delta ZP$ is the zero
point correction from Section 4, and is of order 0.01-0.02 magnitudes,
depending on filter and chip.  (In the initial solution, the $\Delta ZP$
term was omitted, producing a result less than 0.01 magnitudes different
from the one below.  However, to minimize systematic errors, the term
was added and the solution redone after running the CTE solution.)

A robust fit parameter was minimized using the nonlinear minimization
routine $dfpmin$ from Numerical Recipes in C (Press et al. 1992),
providing the best values of the three parameters, and uncertainties
estimated via bootstrap tests.  In the solution, $c_1$ and $c_2$
are the YCTE and XCTE terms, respectively, while $c_0$ is the 
position-independent offset.

For the NGC 2419 data that were used by CM98, 21 stars were found that met
the count and background criteria, producing a $c_0$ value of
$0.02 \pm 0.07$ magnitudes.  Expanding the sample to all 1997 data to
improve the statistics, $c_0$ was determined to be $0.04 \pm 0.05$
magnitudes.  To further improve the statistics, the count range was
changed to $350-700$, producing 114 points and a correction of
$-0.01 \pm 0.02$ magnitudes despite the expected value of 0.12 magnitudes
from the CM98 correction.

Although the uncertainties are significant, a position-independent offset
of 0.23 magnitudes in the faint data is ruled out at almost the 4$\sigma$
level, while the expected correction of 0.12 magnitudes for brighter points
is ruled out at more than the 5$\sigma$ level.  Both samples, however, are
consistent to well within 1$\sigma$ with no anomaly, and therefore the CTE
analysis below does not include a position-independent correction except
for the zero point corrections, $\Delta ZP$.  Further discussion of the
long-short anomaly is given in Section 5.3.

\subsection{CTE Functional Form}

With the result of no position-independent correction, the functional form
of the charge loss will include only XCTE and YCTE corrections.  The first
issue to be determined is whether the charge loss is of the form
\begin{equation}
\Delta counts\ =\ \frac{Y}{800} \hbox{YCTE}\ +\ \frac{X}{800} \hbox{XCTE}
\end{equation}
or
\begin{equation}
\Delta magnitude\ =\ \frac{Y}{800} \hbox{YCTE}\ +\ \frac{X}{800} \hbox{XCTE}
\end{equation}
As is demonstrated in Figure \ref{figFF1}, which shows both magnitude
residuals and the ratio of counts lost as a function of Y for a subset of
the data, the choice of functional form does not make a significant
difference.  However, the plot as a function of magnitude difference gives
the slightly better fit, so the form given in Equation 4 will be used for
this study.
\placefigure{figFF1}

The CTE loss is assumed to depend only on counts, background, temperature,
and epoch.  This assumption was tested by determining the CTE solution for
subsets of the data restricted by chip, gain, and filter.  These subset CTE
determinations produced corrections that were not significantly different
from the full solution, thus verifying the assumption.

Finally, a number of numerical and observational criteria were set out,
which the CTE functional form must satisfy.
\begin{itemize}
\item There must be no combination of counts, background, and epoch
that would cause the formula to become undefined.
\item As the CTE effect is one of lost charge, the correction must never be
negative.
\item The logarithms of the mean magnitude residuals appear to scale
roughly linearly with the logarithms of $counts$ and $background$, implying
a form involving $e^{c_{ct}\ln(counts)}$ and $e^{c_{bg}\ln(background)}$.
\item A sufficiently high background will eliminate the CTE loss.  For
example, stars with less than 1000 $counts$ and more than 1000 $background$
have an average offset of $-0.001\pm0.015$ magnitudes.
\item A sufficiently high count rate will reduce but not eliminate the CTE
loss.  For example, stars with over 21000 $counts$ but under 10 $background$
have an average offset of $0.031\pm0.001$ magnitudes.
\item A sufficiently early time will again reduce but not eliminate the
CTE loss.  For example, the average cold camera offset in 1994 is
$0.020\pm0.001$ magnitudes.
\item High count rates do not eliminate the time dependence, nor do early
epochs eliminate the count dependence.  For example, stars with 21000 or
more $counts$ increase from an average offset of $0.018\pm0.002$ to
$0.041\pm0.004$ magnitudes from 1994 to 2000.  Likewise, the 1994
correction increases from $0.018\pm0.002$ magnitudes for stars with at
least 21000 $counts$ to $0.083\pm0.002$ magnitudes for those between 500
and 1000 $counts$.
\end{itemize}

These observations restrict the set of functional forms that can be used.
Attempts were made to adopt different forms obeying the above restrictions,
with the formulae below producing the best fits to the data.
\begin{equation}
yr=epoch-1996.3
\end{equation}
\begin{equation}
lct=\ln(counts)-7
\end{equation}
\begin{equation}
lbg=\ln(background)-1
\end{equation}
\begin{equation}
\hbox{YCTE}\ =\ \frac{Y}{800}\ [\ y_0\ +\ (y_1+y_2yr) (y_3+e^{-y_4lct}) e^{-y_5lbg-y_6bg}\ ],
\end{equation}
\begin{equation}
\hbox{XCTE}\ =\ \frac{X}{800}\ [\ (x_1+x_2yr) e^{-x_4lct-x_5lbg}\ ].
\end{equation}
The offsets of 1996.3, 7, and 1 were roughly the averages of these values
in the data, and were included to improve numerical stability and to
produce independent coefficients.  (Failing to do so would have provided
correlated coefficients and thus correlated errors.)  The addition of the
$y_6$ parameter was required to eliminate an overcorrection at moderate and
high background levels seen in the first solution.  No similar $x_6$
parameter was needed.  Additionally, $x_0$ and $x_3$ terms analogous to the
$y_0$ and $y_3$ terms were initially included but were determined to be
insignificant (well less than $1\sigma$) and subsequently removed from the
equation.

Note that in the above functional form, all coefficients $y_i$ and $x_i$
must be positive to meet the requirements itemized above. Additionally, to
avoid negative corrections at early epochs (1994.3), $y_1$ must be at
least twice $y_2$ and likewise for $x_1$ and $x_2$.

Separate sets of constants were determined for the warm ($-76^\circ$C) and
cold ($-88^\circ$C) camera observations.  However, all warm camera
observations were made with little or no background and at similar time
(January through April 1994), so no background or time dependence could be
determined.  Additionally, no XCTE loss was measurable in the warm data.

\subsection{Zero Point Corrections}

In addition to the CTE solution, a set of zero point corrections was also
determined.  The first correction was made to search for any offset between
the ground-based Omega Cen and NGC 2419 data.  This correction factor was
determined to be well under 0.01 magnitudes (and less than a $1\sigma$
effect), and was eliminated.

The remaining corrections, as functions of filter, temperature, chip, and
gain, were divided into two sets.  This division, though made primarily for
numerical reasons, was not entirely arbitrary.  The ``ideal'' solution
would naturally involve a zero point correction for every combination of
temperature, filter, chip, and gain, or a total of 272 free parameters.
This total was reduced to 25 by adopting the following assumptions and
simplifications:
\begin{itemize}
\item The principal factors affecting detection efficiency are wavelength
and temperature.  This makes the assumption that the four chips have
similar wavelength-dependent quantum efficiencies (an overall sensitivity
difference would be corrected by item 3 in this list), which was verified
by determining the CTE corrections individually for the four chips.  This
creates 34 filter- and temperature-dependent zero point offsets.
\item Although the non-standard filter offsets are temperature-dependent,
the differences between their corrections and the corrections of the
standard magnitudes of the same color are observed to be independent of
temperature.  This allows the consolidation of the 26 non-standard filter
zero point corrections to 13 corrections relative to the nearest standard
filter, and the reduction of the 34 total temperature/filter zero point
offsets to 21.
\item The principal factors in the readout efficiency are chip and gain
setting, as each of the eight combinations involves slightly different
electronics.  This creates 8 zero point offsets dependent on chip and
gain.  After running the solution, it was determined that the gain
setting made no measurable difference, further reducing these offsets
to 4.
\end{itemize}

Thus the total zero point correction is thus
\begin{equation}
\Delta ZP = \Delta ZP_{T,color} + \Delta ZP_{filter} + \Delta ZP_{chip}.
\end{equation}
As with the CTE loss correction, all of the zero point corrections were
computed in the sense that they should be subtracted from the WFPC2
magnitudes to produce standardized magnitudes.

\section{CTE and Zero Point Solution}

The solution for the CTE effect and zero point corrections was made by
iteration.  The program was a more sophisticated version of that used
to determine the position-independent correction in section 3.1.  In order
to produce an adequate fit parameter, it was necessary to add 0.02
magnitudes (in quadrature) to the uncertainty of every point.  The source
of this additional uncertainty is unknown, and was also noticed by S98.
After the solution, all photometry was re-read, with corrected points
falling more than 0.15 magnitudes and 2$\sigma$ away from the standard
magnitudes removed for the next iteration.  This process continued until
the same points were removed for consecutive iterations.  For the final
iteration, 55910 of the 58728 points were used.

\subsection{CTE Coefficients}

The final coefficients and offsets provide the values to subtract from
the observed WFPC2 magnitudes (calibrated using H95 flight system zero
points, gain ratios, and pixel area ratios) to generate corrected WFPC2
flight system magnitudes.  The CTE coefficients are given (with 68\%
confidence limits) in Tables \ref{tabCTE1} and \ref{tabCTE2} for cold and
warm data, respectively.  As noted above, no XCTE loss was detectable in
the warm camera observations, but can be characterized well in the cold
observations.  The time dependence of the cold XCTE loss is small and is
consistent with the Whitmore (1998) time dependence, while the dependence
on counts is much higher than that on background.  The detection of any
significant background dependence in the XCTE loss is different from S98
and WHC99, but the effect on the corrections is minimal given the size of
the XCTE correction.  The XCTE loss is only a minor effect, with the
worst-case combination in the calibration data ($lct=-1.15$; $bg=1$,
$lbg=-1$; $yr=4$) producing an XCTE ramp of only 0.05 magnitudes.

The YCTE loss is by far the larger of the two.  The primary difference
between warm and cold is the YCTE base value $y_0$, which decreased by 0.08
magnitudes when the instrument was cooled.  As opposed to the XCTE
equation, the YCTE loss is strongly dependent on counts, background, and
time.  Again the time dependence is consistent with the values from
Whitmore (1998).  The worst-case combination will produce a YCTE ramp of
0.50 magnitudes, consistent with the WHC99 result of a $\sim40\%$ loss
for faint stars on low background in early 1999.

A second simple check can be made by determining the corrections for the
conditions used by H95 ($lct=1$; $bg=3$, $lbg=0$; $yr=-1.8$) to determine
their corrections of 0.10 to 0.15 magnitudes in the warm data and 0.04 in
the cold.  These values lead to YCTE ramps of 0.11 magnitudes in the warm
data and 0.03 magnitudes in the cold data, consistent with the H95 CTE
loss estimates.  Detailed comparisons with the recent studies are given in
Section 5.2.
\placetable{tabCTE1}
\placetable{tabCTE2}

\subsection{Zero Point Offsets}

The zero point offsets by color and temperature are given in Table
\ref{tabZP1}.  There is a clear trend of increasing offset with increasing
wavelength for the cold data, and the opposite case in the warm data.
This trend was extrapolated to U (F336W) to compensate for the lack of U
data in the calibration sample.  Because of the extrapolation, the
uncertainties are likely $\sim$0.01 magnitudes.  The cause of these
offsets is unknown, as the H95 calibrations were based on these same set
of observations.  Systematic differences between HSTphot and aperture
photometry are unlikely, especially given the excellent agreement between
the present CTE loss determination and that by WHC99.  It is also not a
time-dependent effect, as this trend is still seen if only the 1994 data
are used.  The easiest explanation would be that H95 used both warm and
cold data, as the offsets would roughly cancel out if added, but this was
not the case.  Compared with the synthetic zero points of Table 28.1 of the
HST data handbook, the new zero points are on average 0.02 magnitudes
fainter, with RMS scatter of 0.02 magnitudes.  The difference is most
likely the result of the handbook zero points, which are only claimed to
be accurate to 0.02 magnitudes.
\placetable{tabZP1}

Table \ref{tabZP2} gives the zero point corrections of the non-standard
filters, which should be subtracted from the observed WFPC2 flight
system magnitudes in order to determine magnitudes on the same system
as the standard filter of the same color.  (In addition, The Table
\ref{tabZP1} value should be subtracted to produce magnitudes corrected
to the ground-based data.)  It is assumed that these values are
nonzero because the H95 transformations for these filters were synthetic
rather than empirical, and thus were not directly tied to any observed
system.  Given the amount of observations made in F450W and F606W, it
is worth noting that while the F450W magnitudes appear to be
correct relative to F439W, F606W magnitudes are off by 0.02 magnitudes
relative to F555W.  A comparison between the new zero points and those
in Table 28.1 of the HST data handbook indicates that the handbook zero
points fare quite poorly, with RMS scatter of 0.08 magnitudes between the
new zero points and those of Table 28.1 for the non-standard filters.  In
comparison, the RMS scatter between the new zero points and the synthetic
zero points of H95 is less than half of that value.
\placetable{tabZP2}

Finally, the chip-to-chip zero point corrections are $0.037\pm0.001$,
$-0.012\pm0.001$, $0.007\pm0.001$, and $0.004\pm0.001$ for PC1, WFC2, WFC3,
and WFC4, respectively.  The presence of non-zero chip-to-chip differences
implies either a minor error in the relative pixel areas reported by H95 or
a sensitivity difference between the four chips.  Although other studies
have determined that chip-to-chip offsets are a function of filter as well
as gain, such an effect is not observed here at any significant level.

\subsection{WFPC2 Calibration Formulae}

Rather than calibrating using H95, and then applying two or three sets of
zero point corrections, it is simpler to apply the corrections to the
calibration process.  The calibration equations, analogous to equations 7
and 8 of H95 but incorporating the pixel area correction, would be
\begin{equation}
\hbox{WFPC2}=-2.5 \log (DN/s) + Z_{FG} + \Delta Z_{CG} - \hbox{CTE}
\end{equation}
and
\begin{equation}
\hbox{SMAG}=-2.5 \log (DN/s) + Z_{FS} + T_1\hbox{SCOL} + T_2\hbox{SCOL}^2 + \Delta Z_{CG} - \hbox{CTE}.
\end{equation}
$\Delta Z_{CG}$ is the zero point modification for chip and gain settings,
and the values can be found in Table \ref{tabCG}.  $Z_{FG}$ in Table
\ref{tabZPall}, and $Z_{FS}$, $T_1$, and $T_2$ in Table \ref{tabXFORM}, and
follow the definitions of H95.  The terms $T_1$ and $T_2$ in Table
\ref{tabXFORM} are reproduced from Tables 7 and 10 of H95.  The CTE
correction should be calculated from Equations 5-9 and Table \ref{tabCTE1}
for cold data or \ref{tabCTE2} for warm data.  Again, it should be noted
that aperture corrections were made to the 0.5 arcsec aperture of H95 and
others, rather than the nominal infinite aperture of the zero points in
Table 28.1 of the HST data handbook.
\placetable{tabCG}
\placetable{tabZPall}
\placetable{tabXFORM}

\section{Tests of the Corrections}

\subsection{Internal Consistency}

Residuals (WFPC2 magnitude minus ground-based magnitude) from before and
after applying the CTE correction and revised calibration are shown in
Figures \ref{figYCTE}-\ref{figBGCTE}, plotted against Y, X, $\ln(counts)$,
and $\ln(background)$, in order to provide a preliminary test of the
correction formulae.  As the figures demonstrate, the correction has
been successful, at least to first order, in reducing the systematic
residuals to under 0.01 magnitudes.
\placefigure{figYCTE}
\placefigure{figXCTE}
\placefigure{figCTCTE}
\placefigure{figBGCTE}

A concern in adopting any functional form for the corrections is that it
will not properly account for second-order factors.  For example, one
frequently-mentioned concern is that the count dependence in the CTE
corrections changes as a function of background level.  This can be readily
tested, with mean residuals for combinations of low and high counts,
background, and epoch shown in Table \ref{tabComp2}.  None of the
residuals in the table are significantly more than 0.005 magnitudes.
\placetable{tabComp2}

However, the most significant source of concern is caused by the fact that
the average calibration data are significantly different from most science
data.  The calibration observations are generally bright stars with little
background (due to short exposure times), while most science observations
are long enough to have a significant background and most projects require
accurate photometry as faint as possible.  The effect of this can be
tested, and the solution appears to have succeeded despite these problems,
with the average residual for points with $background$ between 35 and 105
(typical for most science exposures) being 0.003 magnitudes.

Another potential source of error in the calibration is at the faint end,
as there are insufficient stars in the calibration sample with $<350$
$counts$ (and those that are present naturally have large uncertainties)
to have much effect on the solution.  Therefore, the solution is largely
extrapolated below 350 $counts$ and subject to significant systematic
errors.  Such an effect can be checked by comparing the relative
photometry of the NGC 2419 field, in a manner similar to that of CM98,
but only using the F814W images with no preflash.  The multiple image
version of HSTphot, \textit{multiphot} (Dolphin 2000), was used in order
to reduce the random scatter and force a common object list.  The 100
and 300 second images had sufficient background (at least 1 DN per pixel)
to reduce faint sensitivity and were omitted.  As with the CTE study, the
more crowded chip (WFC2) including the globular cluster was omitted to
improve the photometry.  Figure \ref{fig2419zbg} shows the short (10- and
40-second exposure) magnitudes minus the 1000s magnitudes for these stars
plotted against $\ln(counts)$ scaled from the 1000s image, before CTE
correction in the top panel and after correction in the bottom.  Only
well-fit stars ($\chi \le 2$ and $-0.3 \le sharpness \le 0.3$ are shown.
No systematic error is detectable, aside from the effect of the photometry
cutoff beginning around $\ln(counts)=5.3$, or 200 electrons.  Between
$\ln(counts)$ of 5.3 and 5.5, the median residual after CTE correction is
-0.01 magnitudes, while the median uncorrected residual is 0.15 magnitudes.
However, even down to $\ln(counts)=4.9$ (134 electrons), the mode of the
distribution is within 0.01 magnitudes of zero, giving confidence in the
photometry to the faintest level.
\placefigure{fig2419zbg}

As a final check on the corrections, the corrected WFPC2 data were used
to determine combined magnitudes of the Omega Cen and NGC 2419 standard
stars.  A comparison of ground-based and corrected WFPC2 data is given
in Figure \ref{figSTDS}, showing good agreement.
\placefigure{figSTDS}

\subsection{Comparison with Previous CTE Studies}

In order to make a comparison between the present corrections and those
of S98, WHC99, and SLP00, all four sets of corrections were applied to
the data set used in this study.  To be perfectly fair, it should be noted
that this study's corrections are at somewhat of an advantage, given that
the same data used to determine the corrections are now used for the
comparison.  However, the data set used here is identical (although
larger) than that used in the earlier studies, and the quality of the
photometry should be at least as good as that in the earlier studies given
the use of HSTphot.  Finally, the fact that all major differences are
attributable to obvious causes gives confidence that the comparisons are
accurate.

To eliminate the effects of zero point differences, the present zero point
corrections were applied to the WHC99 study, while the S98 and SLP00 zero
points were applied respectively.  In order to adequately model the SLP00
background dependence, their background parameter was modified for the
observed DoPHOT count-background correlation
\begin{equation}
\hbox{DoPHOT background} = (\hbox{Dark Background}) + 2\times10^{-3} counts.
\end{equation}
Residuals (corrected WFPC2 magnitudes minus standard magnitudes) are given
in Table \ref{tabCompare}.
\placetable{tabCompare}

In terms of the overall corrections in the top line of the table, the
WHC99 equations produce very similar results to the present study.  The
S98 and SLP00 corrections are also reasonably good on average, but
produce considerably more scatter.  The cause of this extra scatter is
clear when examining the remaining lines in Table \ref{tabCompare}:
neither includes a time-dependent term and thus has a trend of increasing
residual with time, in the sense that the 2000 data are significantly
under-corrected in both studies.  Since the WFPC2 magnitude of the average
star in this data set has increased by 0.06 magnitudes from 1994 to 2000
(and that the magnitude of the average star with 350 electrons or less has
increased by 0.18 magnitudes), it would have been surprising had either S98
or SLP00 been consistent with all of the available data.

A plot comparing the present CTE corrections to those of S98 using only
cold F555W and F814W data obtained before 1997 and ignoring zero point
differences is shown in Figure \ref{figCTES98}.  The corrections clearly
agree extremely well for this limited data set, with an average difference
of 0.004 magnitudes and scatter of 0.015 magnitudes.  However, the time
dependence limits the usefulness of the S98 solution, with the 1999-2000
data producing an average difference of 0.061 magnitudes with significantly
more scatter.
\placefigure{figCTES98}

A similar comparison made with SLP00 is shown in Figure \ref{figCTESLP00},
using only F555W and F814W data obtained during 1997.  The effect of the
different functional form used for the two colors is apparent, with the
F555W correction (which has no position-independent term) producing
excellent agreement (under-correcting on average by $0.006\pm0.010$
magnitudes) while the F814W correction (which has a position-independent
term) considerably off (over-correcting on average by $0.026\pm0.025$
magnitudes).  Although the zero point offsets will compensate for this
error in conditions identical to the calibration data used by SLP00,
it is dangerous to assume that this will be the case in general.
\placefigure{figCTESLP00}

Finally, the agreement between the present corrections and those of WHC99,
shown in Figure \ref{figCTEWHC99} for data taken through February 1999, is
excellent.  Although the data show more scatter than Figure \ref{figCTES98},
a figure using the WHC99 corrections restricted to data before 1997 (as
per Figure \ref{figCTES98} would show only 65\% the scatter.  The
differences become significant only at more recent epochs, where the WHC99
functional form (with the count dependence tied to the time dependence)
begins to break down.  For example, the average star with under 1000 counts
observed between February 1998 and February 1999 has an average residual of
$0.10\pm0.14$ magnitudes with the WHC99 corrections, while those of this
study produce an average residual of $0.00\pm0.10$ magnitudes.
\placefigure{figCTEWHC99}

Thus the differences between the results of this study and previous recent
studies can be explained fairly easily.  Neither S98 nor SLP00 included a
time effect, producing large deviations in corrections for stars beyond the
times in which the data were taken.  The remaining significant scatter
between this study and SLP00 is due to the use of a single parameter (the
DoPHOT background level) instead of both the dark background and counts.
Minor differences remain between all four corrections, due to different
choices made in the functional forms.

This comparison of the different CTE solutions appears to explain the
discrepancies, leading to the conclusion that, provided any count
dependence is removed from the background levels, there is no package
dependence in the CTE loss.  It is thus suggested that the present CTE
study, which contains a time dependence and corrects some of the
shortfallings in the WHC99 functional form, should be applicable to all
WFPC2 stellar photometry.

\subsection{The Long-Short Anomaly Revisited}

The previous section shows, reassuringly, that differences between previous
CTE studies are primarily the result of assumptions of time dependence and
the method of background calculation.  However, the different results for
the long-short anomaly (a position-independent charge loss) need to be
reconciled as well.

Positive results regarding the detection of a long-short anomaly have
come from the modified zero points of Saha et al. (1996), Kelson et al.
(1996), and Hill et al. (1998), all of whom found that magnitudes from
long exposures are about 0.05 magnitudes more than those from short
exposures.  CM98 characterized this effect, providing a correction formula
that was based on the number of counts rather than the exposure time.
However, S98 attempted to determine a position-independent loss in his CTE
solution, but found no evidence for it, and Section 3.1 of this paper
likewise finds no evidence.  Finally, Section 5.1, which compares magnitudes
from the 1000-second exposure of NGC 2419 with those from 10- and 40-second
exposures, finds no effect at more than the 0.01 magnitude level.

In order to understand the nature of this effect, a more detailed
comparison between the present data and that of CM98 needs to be made.
Specifically, Figure \ref{fig2419zbg} was taken from the same data used
by CM98 and compared in the same way (short magnitude minus the
1000-second magnitude), and should thus be nearly identical to the 10s
and 40s lines of Figure 11 of CM98 (although CM98 Figure 11 is plotted
against short counts rather than scaled long counts).  However, while CM98
shows a median short minus long error of 0.27 magnitudes for
$\ln(counts)=5.4$, the difference in this study is only -0.01 magnitudes.
The difference between the two studies is also in the pre-CTE correction
photometry, with Figure 9 of CM98 showing a magnitude difference of about
+0.38 magnitudes and the top panel of Figure \ref{fig2419zbg} showing a
difference of +0.15 magnitudes at $\ln(counts)=5.5$, thus ruling out the CTE
solution as the source of the anomaly.

The difference, then, appears to stem from the photometry.  The most
plausible explanation for the apparent long-short anomaly is an
overestimate of the sky by CM98, as well as by previous authors.  The
expected functional form of the magnitude error caused by a sky measurement
error is
\begin{equation}
\Delta mag = -2.5 \log(1-\frac{\pi r^2 \Delta sky}{counts}),
\end{equation}
where $\Delta sky$ is the error per pixel in the sky value and $counts$ is
the number of counts.  For the two-pixel radius used by CM98, a sky error of
+0.58 DN or +4.1 electrons would match the CM98 correction formula to within
0.01 magnitudes for stars with detections of 33 DN or more.  This potential
solution to the problem would also explain the aperture dependence of the
effect seen by CM98, as a larger aperture should be more susceptible to a
sky error.  Finally, this would explain why CM98 found only a count
dependence in the short-long anomaly.  This also would explain the Hill et
al. (1998) remark that the effect appeared to be a constant subtraction of 2
electrons from every star pixel but not the background pixels, as a
background overestimation would do exactly that.

In summary, there appears to be no convincing evidence for a long-short
anomaly.  It would be quite remarkable if a real effect of 0.27 magnitudes
measured by CM98 was reduced to -0.01 magnitudes by an error in HSTphot;
rather it is more likely that an error in the CM98 reduction was responsible
for a false detection of the anomaly.  Given the similarity between the CM98
correction equation and Equation 14 of this paper given $\Delta sky$ error
of 0.58 DN, which agree at the 0.01 magnitude level above 33 DN, as well
as the simple explanation of the aperture dependence of the effect, the
most probable cause of the CM98 result (as well as the other reports of
the long-short anomaly) is an overestimation of the background by a few
electrons.

\section{Summary}

An HSTphot-based CTE and zero point correction study, based on the Omega
Cen and NGC 2419 observations, attempts to accomplish two goals.  First is
a comprehensive testing of HSTphot, which succeeded in reducing well over
1000 WFPC2 images without any problem.  Given the tight agreement between
this study and previous CTE work of S98 and WHC99, it seems that HSTphot is
able to produce photometry without any noticeable systematic effects.

The more ambitious goals of this study were to improve upon existing CTE
determinations, and to determine the cause of differences between the
previously published corrections.  A functional form for the CTE correction
was arrived at semi-empirically, and the coefficients were solved using an
iterative nonlinear minimization process.  The correction formulae were
then applied to the observed data, resulting in no systematic residuals
larger than 0.005 magnitudes.  Additionally, the corrections were applied
to relative photometry of the NGC 2419 field, with no significant errors
found down to the faintest stars measured (under 140 electrons).  New
zero points were also determined for 18 of WFPC2's medium and wide filters,
including empirical calibrations of the non-standard filters, such as F606W.
These new zero points provide evidence of significant errors in the
synthetic zero points given in the HST data handbook.

The corrections were also compared to those from previous studies (S98,
WHC99, and SLP00).  The major differences between the four sets of
corrections are understandable in terms of the lack of a time dependence
in S98 or SLP00 and the lack of a count dependence in SLP00.  The primary
difference between this study and WHC99 is the use of an improved
characterization of the count and time dependencies.  In terms of producing
smaller systematic and random residuals over the full time baseline, the
present set of corrections proved superior to all three other
formulations, although the WHC99 CTE equations will also produce
corrections accurate to a few hundredths of a magnitude in all cases except
for recent (1998 and later) data with low counts.  The previous detections
and characterizations of the long-short error are believed to result from
sky overestimation, an error which would cause the count dependence
characterized by CM98 and fit their correction formula to within 0.01
magnitudes, as well as explaining the aperture dependence and lack of
background dependence.

It is concluded that, given the understanding of the differences between
this and other CTE studies, the present corrections should be valid for use
on all WFPC2 stellar photometry, regardless of the photometry procedure,
and are an improvement on previous work.  However, due to differences in
the determination of background (especially by DoPHOT), the contribution of
starlight to the background level must be fit and removed in order to
determine the dark background for a star.

\acknowledgments

I would like to thank Alistair Walker and Peter Stetson for providing the
ground-based Omega Cen and NGC 2419 photometry, respectively.
This work was supported by NASA through grants GO-02227.06-A and GO-07496
from Space Telescope Science Institute.

\clearpage
\begin{figure}
\caption{Charge loss and magnitude loss as a function of Y for a subset of the data.  The heavy dots are the average values. \label{figFF1}}
\end{figure}

\begin{figure}
\caption{CTE effect before (above) and after (below) correction, plotted against Y \label{figYCTE}}
\end{figure}

\begin{figure}
\caption{CTE effect before (above) and after (below) correction, plotted against X \label{figXCTE}}
\end{figure}

\begin{figure}
\caption{CTE effect before (above) and after (below) correction, plotted against $\ln(counts)$ \label{figCTCTE}}
\end{figure}

\begin{figure}
\caption{CTE effect before (above) and after (below) correction, plotted against $\ln(background)$ \label{figBGCTE}}
\end{figure}

\begin{figure}
\caption{NGC 2419 F814W relative photometry, with short minus long magnitudes for zero preflash 10- and 40-second images plotted against the number of counts. \label{fig2419zbg}}
\end{figure}

\begin{figure}
\caption{Omega Centauri and NGC 2419 photometry, with ground-based data in (a) and (c) and calibrated HSTphot data in (b) and (d) \label{figSTDS}}
\end{figure}

\begin{figure}
\caption{Differences in CTE and zero point corrections between S98 and the present study for cold F555W and F814W data obtained before 1997.  Positive values mean that the S98 correction is larger, thus producing a smaller (brighter) corrected magnitude. \label{figCTES98}}
\end{figure}

\begin{figure}
\caption{Same as Figure \ref{figCTES98} for data obtained in 1997 and 1998, compared with SLP00 \label{figCTESLP00}}
\end{figure}

\begin{figure}
\caption{Same as Figure \ref{figCTES98} for data cold obtained through February 1999, compared with WHC99 \label{figCTEWHC99}}
\end{figure}

\clearpage
\begin{deluxetable}{lrl}
\tablecaption{Cold CTE Coefficients \label{tabCTE1}}
\tablehead{
\colhead{$c_i$} &
\colhead{Value} &
\colhead{Description}}
\startdata
$y_{0}$  & $ 0.018 \pm 0.003$ &  YCTE base \nl
$y_{1}$  & $ 0.097 \pm 0.005$ &  YCTE time-independent \nl
$y_{2}$  & $ 0.041 \pm 0.002$ &  YCTE time-dependent \nl
$y_{3}$  & $ 0.088 \pm 0.031$ &  YCTE count-independent \nl
$y_{4}$  & $ 0.507 \pm 0.019$ &  YCTE count-dependent \nl
$y_{5}$  & $ 0.035 \pm 0.025$ &  YCTE background-dependent \nl
$y_{6}$  & $ 0.042 \pm 0.008$ &  YCTE background-dependent $e^{-bg}$ \nl
$x_{1}$  & $ 0.024 \pm 0.002$ &  XCTE time-independent \nl
$x_{2}$  & $ 0.002 \pm 0.001$ &  XCTE time-dependent \nl
$x_{4}$  & $ 0.196 \pm 0.042$ &  XCTE count-dependent \nl
$x_{5}$  & $ 0.126 \pm 0.034$ &  XCTE background-dependent \nl
\enddata
\end{deluxetable}

\clearpage
\begin{deluxetable}{lrl}
\tablecaption{Warm CTE Coefficients \label{tabCTE2}}
\tablehead{
\colhead{$c_i$} &
\colhead{Value} &
\colhead{Description}}
\startdata
$y_{0}$  & $ 0.103 \pm 0.003$ &  YCTE base \nl
$y_{3}$  & $ 0.028 \pm 0.004$ &  YCTE count-dependent constant\nl
$y_{4}$  & $ 0.959 \pm 0.089$ &  YCTE count dependence \nl
\enddata
\end{deluxetable}

\clearpage
\begin{deluxetable}{lrr}
\tablecaption{Color and Temperature Corrections \label{tabZP1}}
\tablehead{
\colhead{Color} &
\colhead{Cold $\Delta ZP_{color}$} &
\colhead{Warm $\Delta ZP_{color}$}}
\startdata
U\tablenotemark{1} & $-0.023 \pm 0.010$ & $ 0.045 \pm 0.010$\nl
B                  & $-0.016 \pm 0.001$ & $ 0.034 \pm 0.002$\nl
V                  & $-0.009 \pm 0.001$ & $ 0.013 \pm 0.001$\nl
R                  & $-0.007 \pm 0.001$ & $ 0.013 \pm 0.002$\nl
I                  & $ 0.012 \pm 0.001$ & $-0.014 \pm 0.001$\nl
\enddata
\tablenotetext{1}{U zero point corrections are extrapolated from the
other four colors.}
\end{deluxetable}

\clearpage
\begin{deluxetable}{lrr}
\tablecaption{Zero Point Corrections for Non-Standard Filters \label{tabZP2}}
\tablehead{
\colhead{Filter} &
\colhead{Standard} &
\colhead{$\Delta ZP_{filter}$}}
\startdata
F380W  & F336W & $-0.084 \pm 0.004$\nl
F410M  & F439W & $-0.055 \pm 0.006$\nl
F450W  & F439W & $ 0.006 \pm 0.004$\nl
F467M  & F439W & $ 0.062 \pm 0.004$\nl
F547M  & F555W & $ 0.005 \pm 0.003$\nl
F569W  & F555W & $ 0.017 \pm 0.002$\nl
F606W  & F555W & $ 0.019 \pm 0.002$\nl
F622W  & F675W & $ 0.006 \pm 0.007$\nl
F702W  & F675W & $-0.010 \pm 0.002$\nl
F785LP & F814W & $-0.008 \pm 0.003$\nl
F791W  & F814W & $ 0.026 \pm 0.005$\nl
F850LP & F814W & $ 0.001 \pm 0.005$\nl
F1042M & F814W & $ 0.004 \pm 0.006$\nl
\enddata
\end{deluxetable}

\clearpage
\begin{deluxetable}{lrr}
\tablecaption{$\Delta Z_{CG}$ Values \label{tabCG}}
\tablehead{
\colhead{Chip} &
\colhead{gain=14} &
\colhead{gain=7}}
\startdata
PC1  & $-0.044\pm0.001$ & $ 0.701\pm0.001$ \nl
WFC2 & $ 0.007\pm0.000$ & $ 0.761\pm0.000$ \nl
WFC3 & $-0.007\pm0.000$ & $ 0.749\pm0.000$ \nl
WFC4 & $-0.005\pm0.000$ & $ 0.722\pm0.000$ \nl
\enddata
\end{deluxetable}

\clearpage
\begin{deluxetable}{lrr}
\tablecaption{New Flight System Zero Points \label{tabZPall}}
\tablehead{
\colhead{Filter} &
\colhead{Cold $Z_{FG}$} &
\colhead{Warm $Z_{FG}$}}
\startdata
F336W  & $18.528 \pm 0.010$ & $18.460 \pm 0.010$\nl
F380W  & $20.218 \pm 0.004$ & $20.169 \pm 0.005$\nl
F410M  & $18.576 \pm 0.007$ & $18.527 \pm 0.007$\nl
F439W  & $20.086 \pm 0.001$ & $20.036 \pm 0.002$\nl
F450W  & $21.183 \pm 0.004$ & $21.133 \pm 0.005$\nl
F467M  & $19.114 \pm 0.004$ & $19.065 \pm 0.004$\nl
F547M  & $20.843 \pm 0.003$ & $20.820 \pm 0.003$\nl
F555W  & $21.734 \pm 0.001$ & $21.712 \pm 0.001$\nl
F569W  & $21.411 \pm 0.003$ & $21.389 \pm 0.003$\nl
F606W  & $22.075 \pm 0.002$ & $22.052 \pm 0.002$\nl
F622W  & $21.544 \pm 0.007$ & $21.525 \pm 0.007$\nl
F675W  & $21.241 \pm 0.001$ & $21.221 \pm 0.002$\nl
F702W  & $21.645 \pm 0.002$ & $21.626 \pm 0.003$\nl
F785LP & $19.873 \pm 0.003$ & $19.899 \pm 0.003$\nl
F791W  & $20.669 \pm 0.005$ & $20.695 \pm 0.005$\nl
F814W  & $20.827 \pm 0.001$ & $20.853 \pm 0.001$\nl
F850LP & $19.126 \pm 0.005$ & $19.153 \pm 0.005$\nl
F1042M & $15.351 \pm 0.006$ & $15.378 \pm 0.006$\nl
\enddata
\end{deluxetable}

\clearpage
\begin{deluxetable}{lccrrrrrr}
\tablecaption{New Transformations \label{tabXFORM}}
\tablehead{
\colhead{Filter} &
\colhead{SMAG} &
\colhead{SCOL} &
\colhead{$T_1$} &
\colhead{$T_2$} &
\colhead{$Z_{FS,cold}$} &
\colhead{$Z_{FS,warm}$} &
\colhead{$C_{min}$} &
\colhead{$C_{max}$}}
\startdata
F300W  & U & (U-B) & -1.532 & -0.519 & 18.156 & 18.144 &      & -0.2 \nl
F300W  & U & (U-B) & -0.427 &  0.138 & 18.181 & 18.169 &  0.2 &  1.0 \nl
F336W  & U & (U-B) & -0.844 & -0.160 & 18.528 & 18.460 &      &      \nl
F336W  & U & (U-V) & -0.240 &  0.048 & 18.787 & 18.719 &      &      \nl
F336W  & U & (U-R) & -0.172 &  0.041 & 18.820 & 18.752 &      &      \nl
F336W  & U & (U-I) & -0.149 &  0.038 & 18.840 & 18.772 &      &      \nl
F380W  & B & (B-V) & -0.581 &  0.777 & 20.243 & 20.194 &      &  0.5 \nl
F380W  & B & (B-V) & -0.943 &  0.103 & 20.595 & 20.546 &  0.5 &  1.4 \nl
F410M  & B & (B-V) & -0.183 & -0.287 & 18.886 & 18.837 &      &  1.4 \nl
F439W  & B & (U-B) & -0.103 & -0.046 & 20.073 & 20.023 &      &      \nl
F439W  & B & (B-V) &  0.003 & -0.088 & 20.086 & 20.036 &      &      \nl
F439W  & B & (B-R) &  0.019 & -0.049 & 20.080 & 20.030 &      &      \nl
F439W  & B & (B-I) &  0.005 & -0.023 & 20.083 & 20.033 &      &      \nl
F450W  & B & (B-V) &  0.230 & -0.003 & 21.185 & 21.135 &      &  1.4 \nl
F467M  & B & (B-V) &  0.480 & -0.299 & 19.121 & 19.072 &      &  0.5 \nl
F467M  & B & (B-V) &  0.432 & -0.002 & 19.072 & 19.023 &  0.5 &  1.4 \nl
F547M  & V & (V-I) &  0.027 & -0.032 & 20.838 & 20.815 &      &  1.1 \nl
F547M  & V & (V-I) &  0.049 & -0.013 & 20.790 & 20.767 &  1.1 &      \nl
F555W  & V & (U-V) & -0.014 &  0.005 & 21.715 & 21.693 &      &      \nl
F555W  & V & (B-V) & -0.060 &  0.033 & 21.734 & 21.712 &      &      \nl
F555W  & V & (V-R) & -0.121 &  0.120 & 21.739 & 21.717 &      &      \nl
F555W  & V & (V-I) & -0.052 &  0.027 & 21.734 & 21.712 &      &      \nl
F569W  & V & (V-I) &  0.089 & -0.003 & 21.409 & 21.387 &      &  2.0 \nl
F569W  & V & (V-I) & -0.125 &  0.022 & 21.741 & 21.719 &  2.0 &      \nl
F606W  & V & (V-I) &  0.254 &  0.012 & 22.084 & 22.061 &      &  2.0 \nl
F606W  & V & (V-I) & -0.247 &  0.065 & 22.874 & 22.851 &  2.0 &      \nl
F622W  & R & (V-R) & -0.252 & -0.111 & 21.558 & 21.539 &      &      \nl
F675W  & R & (U-R) &  0.039 & -0.007 & 21.261 & 21.241 &      &      \nl
F675W  & R & (B-R) &  0.092 & -0.017 & 21.242 & 21.222 &      &      \nl
F675W  & R & (V-R) &  0.253 & -0.125 & 21.241 & 21.221 &      &      \nl
F675W  & R & (R-I) &  0.273 & -0.066 & 21.232 & 21.212 &      &      \nl
F702W  & R & (V-R) &  0.343 & -0.177 & 21.650 & 21.631 &      &  0.6 \nl
F702W  & R & (V-R) &  0.486 & -0.079 & 21.528 & 21.509 &  0.6 &      \nl
F785LP & I & (V-I) &  0.091 &  0.020 & 19.876 & 19.902 &      &      \nl
F791W  & I & (V-I) & -0.029 & -0.004 & 20.669 & 20.695 &      &  1.0 \nl
F791W  & I & (V-I) & -0.084 &  0.011 & 20.710 & 20.736 &  1.0 &      \nl
F814W  & I & (U-I) & -0.018 &  0.002 & 20.803 & 20.829 &      &      \nl
F814W  & I & (B-I) & -0.031 &  0.007 & 20.823 & 20.849 &      &      \nl
F814W  & I & (V-I) & -0.062 &  0.025 & 20.827 & 20.853 &      &      \nl
F814W  & I & (R-I) & -0.112 &  0.084 & 20.827 & 20.853 &      &      \nl
F850LP & I & (V-I) &  0.160 &  0.023 & 19.108 & 19.135 &      &      \nl
F1042M & I & (V-I) &  0.350 &  0.022 & 15.298 & 15.325 &      &      \nl
\enddata
\end{deluxetable}

\clearpage
\begin{deluxetable}{lrrrr}
\tablecaption{Second-Order Residuals after Correction \label{tabComp2}}
\tablehead{
\colhead{} &
\colhead{$bg< 100$} &
\colhead{$bg> 100$} &
\colhead{$ct<2000$} &
\colhead{$ct>2000$}}
\startdata
$ct<2000$ & $ 0.003\pm0.001$ & $-0.003\pm0.005$ & & \nl
$ct>2000$ & $-0.001\pm0.001$ & $-0.008\pm0.003$ & & \nl
$yr<1996$ & $ 0.001\pm0.001$ & $-0.010\pm0.007$ & $ 0.004\pm0.001$ & $-0.001\pm0.001$ \nl
$yr>1996$ & $ 0.000\pm0.001$ & $-0.006\pm0.002$ & $ 0.001\pm0.001$ & $-0.001\pm0.001$ \nl
\enddata
\end{deluxetable}

\clearpage
\begin{deluxetable}{rrrrrr}
\tablecaption{Residuals From Four CTE and Zero Point Correction Systems \label{tabCompare}}
\tablehead{
\colhead{Year} &
\colhead{Npoints} &
\colhead{S98} &
\colhead{WHC99} &
\colhead{SLP00} &
\colhead{Present Work}}
\startdata
all & 28221 & $ 0.031\pm0.103$ & $-0.009\pm0.092$ & $ 0.010\pm0.099$ & $-0.001\pm0.089$ \nl
1994 & 6504 & $ 0.010\pm0.105$ & $-0.008\pm0.106$ & $-0.018\pm0.108$ & $-0.001\pm0.104$ \nl
1995 & 6917 & $ 0.019\pm0.097$ & $-0.003\pm0.092$ & $-0.011\pm0.093$ & $ 0.003\pm0.089$ \nl
1996 & 2003 & $ 0.015\pm0.069$ & $-0.018\pm0.064$ & $ 0.005\pm0.067$ & $-0.005\pm0.064$ \nl
1997 & 3436 & $ 0.030\pm0.085$ & $-0.008\pm0.080$ & $ 0.029\pm0.083$ & $ 0.002\pm0.079$ \nl
1998 & 1956 & $ 0.044\pm0.092$ & $-0.014\pm0.085$ & $ 0.022\pm0.087$ & $-0.007\pm0.082$ \nl
1999 & 4259 & $ 0.066\pm0.112$ & $-0.008\pm0.096$ & $ 0.040\pm0.103$ & $ 0.002\pm0.089$ \nl
2000 & 3146 & $ 0.065\pm0.105$ & $-0.018\pm0.089$ & $ 0.040\pm0.096$ & $-0.004\pm0.082$ \nl
\enddata
\end{deluxetable}

\end{document}